\documentclass[a4paper,12pt]{article}
\usepackage{amssymb,amsthm,latexsym}

\newcommand{\med}[1]{\langle #1 \rangle}
\newtheorem{theorem}{Theorem}

\title{SPIN GLASSES}

\author{
Francesco Guerra\footnote{\
e-mail: {\tt francesco.guerra@roma1.infn.it}} \\
{\small {\itshape Dipartimento di Fisica, Universit\`a di Roma ``La Sapienza''}}
\\
{\small {\itshape INFN, Sezione di Roma1, Piazzale A. Moro 2, 00185 Roma, 
Italy}}
} 
\date{\today}
\begin{document}
\maketitle

\section{Introduction}

From a physical point of view, spin glasses, as dilute magnetic alloys, are very interesting systems. 
They are characterized by such features as exhibiting a new magnetic 
phase, where magnetic moments are frozen into disordered equilibrium 
orientations, without any long-range order. See for example \cite{Y} for 
general reviews, and also \cite{stein} 
for a very readable account about the physical properties of spin glasses.
The experimental laboratory study of spin glasses is a very difficult 
subject, because of their peculiar properties. In particular the existence 
of very slowly relaxing modes, with consequent memory effects, makes 
difficult to realize the very basic physical concept of a system at 
thermodynamical 
equilibrium, at a given temperature.

From a theoretical point of view some models have been proposed, which try 
to capture the essential physical features of spin glasses, in the frame 
of very simple assumptions.

The basic model has been proposed by Edwards and Anderson \cite{EA} many years 
ago. It 
is a simple extension of the well known nearest neighbour Ising model. On a large 
region $\Lambda$ of the unit lattice in $d$ dimensions, we associate an Ising spin 
$\sigma(n)$ to each lattice site $n$, and then we introduce a lattice 
Hamiltonian
\begin{equation}
\label{HEA}
H_{\Lambda}(\sigma, J) = -\sum_{(n,n^{\prime})} J(n, n^{\prime}) \sigma(n) 
\sigma(n^{\prime}).
\end{equation}
Here, the sum runs over all couples of nearest neighbour sites in 
$\Lambda$, and $J$ are quenched random couplings, assumed for simplicity to 
be independent identically distributed random variables, with centered unit 
Gaussian distribution. The quenched character of the $J$ means that they 
do not partecipate to thermodynamic equilibrium, but act as a kind of 
random external noise on the coupling of the $\sigma$ variables. In the 
expression of the Hamiltonian, 
we have indicated with $\sigma$ the set of all $\sigma(n)$, and with $J$ 
the set of all $J(n, n^{\prime})$. The region $\Lambda$ must be taken very 
large, by letting it invade all lattice in the limit. The physical motivation 
for this choice is that for real spin glasses the
interaction between the spins dissolved in the matrix of the alloy oscillates 
in sign according to distance. This effect is taken into account in the 
model through the random character of the couplings between spins.

Even though very drastic simplifications have been introduced in the 
formulation of this model, as compared to the extremely more complicated 
nature of physical spin glasses, nevertheless a rigorous study of all 
properties emerging from the static and dynamic behavior of a thermodynamic 
system of this kind is far from beeing complete. In particular, with 
reference to static equilibrium properties, it is not possible yet to 
reach a completely substantiated description of the phases emerging in the 
low temperature region. Even physical intuition gives completely different 
guesses for different people.

In the same way as a mean field version can be associated to the ordinary 
Ising model, so it is possible for the disordered model described by 
(\ref{HEA}). Now we consider a number of sites $i=1,2,\dots,N$, and let each 
spin $\sigma(i)$ at site $i$ interact with all other spins, with the 
intervention of a quenched noise $J_{ij}$. The precise form of the 
Hamiltonian will be given in the following.      

This is the mean field model for spin glasses, 
introduced by David Sherrington and 
Scott Kirkpatrick more that thirty years ago \cite{SK}, \cite{KS}. It is a 
celebrated model. 
Hundreds and hundreds of articles have been devoted to its study during the 
years, appearing in the theoretical physics literature.

The relevance of the model stems surely from the fact that it is intended 
to represent some important features of the physical spin glass systems, 
of great interest for their peculiar properties, at least at the level of 
the mean field approximation. 

But another important source of interest is connected with the fact that 
disordered systems, of the Sherrington-Kirkpatrick type, and their 
generalizations, seems to play a very important role for theoretical and 
practical assessments about hard optimization problems, as it is shown for 
example by Mark M\'ezard, Giorgio Parisi and Riccardo Zecchina in \cite{MPZ}.

It is interesting to remark that the original paper was entitled 
``Solvable Model of a Spin-Glass'', while a previous draft, as told
by David Sherrington, contained even the stronger denomination ``Exactly 
Solvable''. However, it turned out that the very natural solution devised 
by the authors is valid only at high temperatures, or for large external 
magnetic fields. While, at low temperatures, the proposed solution 
exhibits a nonphysical drawback given by a negative entropy, as properly 
recognized by the authors in their very first paper.

It took some years to find an acceptable solution. This was done by Giorgio 
Parisi in a series of papers, by marking a radical departure from the 
previous methods. In fact, a very deep method of ``spontaneous replica 
symmetry breaking'' was developed. As a consequence the physical content of 
the theory was encoded in a functional order parameter of new type, and a 
remarkable structure emerged for the pure states of the theory, a kind of 
hierarchical, ultrametric organization. These very interesting 
developments, due to Giorgio Parisi, and his coworkers, are explained in a 
brilliant way in the classical book \cite{MPV}. Part of this structure will be 
recalled in the following.

It is important to remark that Parisi solution is presented in the form of 
an ingenious and clever \textit{Ansatz}. Until few years ago it was not known 
whether this \textit{Ansatz} would give the true solution for the model, in 
the so called thermodynamic limit, when the size of the system becomes 
infinite, or it would be only a very good approximation for the true 
solution.

The general structures offered by the Parisi solution, and their possible 
generalizations for similar models, exhibit an extremely rich and 
interesting mathematical content. Very appropriately, Michel Talagrand has 
inserted a strongly suggestive sentence in the title to his recent book 
\cite{T}: ``Spin glasses: a challenge for mathematicians''.

As a matter of fact, how to face this challenge is a very difficult 
problem. Here we would like to recall the main features of a very 
powerful method, yet extremely simple in its very essence, based on a 
comparison and interpolation argument on sets of Gaussian random variables.

The method found its first simple application in \cite{Gsum}, where it was 
shown that the Sherrington-Kirkpatrick replica symmetric approximate 
solution was a rigourous lower bound for the quenched free energy of the 
system, uniformly in the size. Then, it was possible to reach a long 
waited result \cite{GTthermo}: the convergence of the free energy density in the 
thermodynamic limit, by an intermediate step where the quenched free 
energy was shown to be subadditive in the size of the system.

Moreover, still by interpolation on families of Gaussian random variables, 
the first mentioned result was extended to give a rigorous proof that the 
expression given by the Parisi \textit{Ansatz} is also a lower bound for 
the quenched free energy of the system, uniformly in the size \cite{Grepli}. 
The method gives not only the bound, but also the explicit form of the 
correction in a quite involved form. As a recent and very important result, 
along the task of facing the challenge, Michel Talagrand has been able to 
dominate these correction terms, showing that they vanish in the 
thermodynamic limit. This milestone achievement was firstly announced in a 
short note \cite{Tb}, containing only a synthetic sketch of the proof, and then 
presented with all details in a long paper to be published on Annals of 
Mathematics \cite{Topus}.

The interpolation method is also at the basis of the far reaching 
generalized variational principle proven by Michel Aizenman, Robert Sims and 
Shannon Starr in \cite{ASS}.

In our presentation, we will try to be as self-contained as possible. We 
will give all definitions, explain the basic structure of the interpolation 
method, and show how some of the results are obtained. We will concentrate 
mostly on questions connected with the free energy, its properties of 
subadditivity, the existence of the infinite volume limit, and the replica 
bounds.

For the sake of comparison, and in order to provide a kind of warm up, we 
will recall also some features of the standard elementary mean field model 
of ferromagnetism, the so called Curie-Weiss model. We will concentrate 
also here on the free energy, and systematically exploit elementary 
comparison and interpolation arguments. This will show the strict analogy 
between the treatment of the ferromagnetic model and the developments in 
the mean field spin glass case. Basic roles will be played in the two 
cases, but with different expressions, by positivity and convexity 
properties.

Then, we will consider the problem of connecting results for the mean 
field case to the short range case. An intermediate position is occupied by 
the so called diluted models. They can be studied through a  
generalization of the methods exploited in the mean field case, as shown 
for example in \cite{LDS}. 

The organization of the paper is as follows. In Section 2, we introduce 
the ferromagnetic model and discuss behavior and properties of the free 
energy in the thermodynamic limit, by emphasing, in this very elementary 
case, the comparison and interpolation methods that will be also 
exploited, in a different context, in the spin glass case. 

Section 3 is devoted to the basic features of the mean field spin glass 
models, by introducing all necessary definitions.

In Section 4, we introduce, for generic Gaussian interactions, some 
important 
formulae, concerning the derivation with respect to the strength of the 
interaction, and the Gaussian comparison and interpolation 
method. 

In next Section 5 we give simple applications to the mean field spin glass 
model, in particular to the existence of the infinite 
volume limit of the quenched free energy \cite{GTthermo}, and to the proof of general 
variational bounds, by following the useful strategy developed in \cite{ASS}.

Section 6 will briefly recall the main features of the Parisi 
representation, and will state the main theorem concerning the free energy.

In Section 7 we will make some mention about results for diluted models.

Finally, in Section 8, we attack the problem of connecting the results for 
the mean field case to the more realistic short range models.

Section 9 will be devoted to conclusions and outlook for future 
foreseen developments.

Our treatment will be as simple as possible, by relying on the basic 
structural properties, and by describing methods of presumably very long 
lasting power. The enphasis given to the mean field case reflects the 
status of research. May be that after some years from now this review would  
be written according to completely different patterns.

\section {A warm up. The mean field ferromagnetic model. Structure and results.}

The mean field ferromagnetic model is among the simplest models of 
statistical mechanics. However, it contains very interesting features, in 
particular a phase transition, characterized by spontaneous magnetization, 
at low temperatures. 
We refer to standard textbooks, for example \cite{Stanley}, for a full 
treatment, and a complete appreciation of the model in the frame of the 
theory of ferromagnetism. Here we consider firstly some properties of the free 
energy, easily obtained through comparison methods.

The generic configuration of the  mean field ferromagnetic model is defined 
through Ising spin variables
$\sigma_{i}=\pm 1$,  attached to each site $i=1,2,\dots,N$.

The Hamiltonian of the model, in some external field of strength $h$,  
is given by the mean field expression
\begin{equation}\label{H}
H_N(\sigma,h)=-{1\over N}\sum_{(i,j)}\sigma_i\sigma_j
-h\sum_{i}\sigma_i.
\end{equation}
Here, the first sum extends to all $N(N-1)/2$ site couples, 
and the second to all sites.

For a given inverse temperature $\beta$, let us now introduce the 
partition function $Z_{N}(\beta,h)$ and 
the free energy per site
$f_{N}(\beta,h)$,  
according to the well known definitions
\begin{eqnarray}\label{Z}
&&Z_N(\beta,h)=\sum_{\sigma_1\dots\sigma_N}\exp(-\beta H_N(\sigma,h)),\\
\label{f}
&&-\beta f_N(\beta,h)=N^{-1} E\log Z_N(\beta,h).
\end{eqnarray}

It is also convenient to define the average spin magnetization
\begin{equation}\label{m}
m={1\over N}\sum_{i}\sigma_i.
\end{equation}

Then, it is immediately seen that the Hamiltonian in (\ref{H}) can be 
equivalently written as
\begin{equation}\label{Hm}
H_N(\sigma,h)=-{1\over2} N m^2
-h\sum_{i}\sigma_i,
\end{equation}
where an unessential constant term has been neglected. In fact we have
\begin{equation}\label{HHm}
\sum_{(i,j)}\sigma_i\sigma_j = {1\over2} \sum_{i,j;i\ne j}\sigma_i\sigma_j=
{1\over2} N^2 m^2 - {1\over2} N, 
\end{equation}
where the sum over all couples has been equivalently written as one half the sum 
over all $i,j$ with $i\ne j$, and the diagonal terms with $i=j$ have been added 
and 
subtracted out. Notice that they give a constant because $\sigma_i^2=1$.

Therefore, the partition function in (\ref{Z}) can be equivalently substituted 
by the expression
\begin{equation}\label{Z'}
Z_N(\beta,h)=\sum_{\sigma_1\dots\sigma_N}\exp({1\over2} \beta N m^2)
\exp(\beta h \sum_{i}\sigma_i), 
\end{equation} 
which will be our starting point.

Our interest will be in the $\lim_{N\to\infty} N^{-1} \log Z_N(\beta,h)$.
To this purpose, let us establish the important subadditivity property, 
holding for the splitting of the big N site system in two smaller $N_1$ 
site and $N_2$ site systems, respectively, with $N=N_1+N_2$,
\begin{equation}\label{sub}
\log Z_N(\beta,h)\le \log Z_{N_1}(\beta,h) + \log Z_{N_2}(\beta,h).
\end{equation}
The proof is very simple. Let us denote, in the most natural way, by 
$\sigma_1,\dots,\sigma_{N_1}$ the spin variables for the first subsystem, 
and by $\sigma_{N_1 +1},\dots,\sigma_{N}$ the $N_2$ spin variables of the 
second subsystem. Introduce also the subsystem magnetizations $m_1$ and 
$m_2$, by adapting the definition (\ref{m}) to the smaller systems, in such 
a way that
\begin{equation}\label{m'}
N m= N_1 m_1 + N_2 m_2.
\end{equation}
Therefore, we see that the large system magnetization $m$ is the linear 
convex combination of the smaller system ones, according to the obvious
\begin{equation}\label{m''}
m= {N_1\over N} m_1 + {N_2\over N} m_2.
\end{equation}
Since the mapping $m\to m^2$ is convex, we have also the general bound, 
holding for all values of the $\sigma$ variables
\begin{equation}\label{m2}
m^2 \le {N_1\over N} m_1^2 + {N_2\over N} m_2^2.
\end{equation}
Then, it is enough to substitute the inequality in the definition 
(\ref{Z'}) of 
$Z_N(\beta,h)$, and recognize that we achieve factorization with 
respect to the two subsystems, and therefore the inequality 
$Z_N\le Z_{N_1} Z_{N_2}$. So we have established (\ref{sub}). From 
subadditivity, the existence of the limit follows by a simple argument, as 
explained for example in \cite{ruelle}. In fact, we have
\begin{equation}\label{lim}
\lim_{N\to\infty}N^{-1}\log Z_N(\beta,h)= \inf_{N}N^{-1}\log Z_N(\beta,h).
\end{equation}
Now we will calculate explicitely this limit, by introducing an order 
parameter $M$, a trial function, and an appropriate variational scheme.
In order to get a lower bound, we start from the elementary inequality 
$m^2 \ge 2 m M - M^2$, holding for any value of $m$ and $M$. By inserting 
the inequality in the definition (\ref{Z'}) we arrive at a factorization of 
the sum over $\sigma$'s. The sum can be explicitely calculated, and we 
arrive immediately to the lower bound, uniform in the size of the system,
\begin{equation}\label{lb}
N^{-1}\log Z_N(\beta,h) \ge \log2 + \log\cosh\beta(h+M) - {1\over2}\beta 
M^2,
\end{equation}
holding for any value of the trial order parameter $M$. Clearly it is 
convenient to take the supremum over $M$. Then we establish the optimal 
uniform lower bound
\begin{equation}\label{lb'}
N^{-1}\log Z_N(\beta,h) \ge \sup_M(\log2 + \log\cosh\beta(h+M) - {1\over2}\beta 
M^2).
\end{equation}

It is simple to realize that the supremum coincides with the limit as 
$N\to\infty$. To this purpose we follow the following simple 
procedure. Let us consider all possible values of the variable $m$. There 
are $N+1$ of them, corresponding to any number $K$ of possible spin flips, 
starting from a given $\sigma$ configuration, $K=0,1,\dots,N$. Let us 
consider the trivial decomposition of the identity, holding for any $m$,
\begin{equation}\label{sum}  
1=\sum_M \delta_{m M},
\end{equation}
where $M$ in the sum runs over the $N+1$ possible values of $m$, and 
$\delta$ is Kroneker delta, beeing equal to $1$ if $M=N$, and zero 
otherwise. Let us now insert (\ref{sum}) in the definition (\ref{Z'}) of the 
partition function inside the sum over $\sigma$'s, and invert the two sums. 
Because of the forcing 
$m=M$ given by the $\delta$, we can write $m^2 = 2 m M - M^2$ inside the 
sum. Then if we neglect the $\delta$, by using the trivial $\delta \le 1$, 
we have un upper bound, where the sum over $\sigma$'s can be explicitily 
performed as before. Then it is enough to take the upper bound with 
respect to $M$, and consider that there are $N+1$ terms in the now trivial 
sum over $M$, in order to arrive at the upper bound
\begin{equation}\label{ub}
N^{-1}\log Z_N(\beta,h) \le \sup_M(\log2 + \log\cosh\beta(h+M) - {1\over2}\beta 
M^2) + N^{-1}\log (N+1).
\end{equation}
Therefore, by going to the limit as $N\to\infty$, we can collect all our 
results in the form of the following theorem giving the full 
characterization of the thermodynamic limit of the free energy.
\begin{theorem}
\label{tlim}
For the mean field ferromagnetic model we have
\begin{eqnarray}    
&&\lim_{N\to\infty}N^{-1}\log Z_N(\beta,h) =  \inf_N N^{-1}\log Z_N(\beta,h)\\
&&=\sup_M(\log2 + \log\cosh\beta(h+M) - {1\over2}\beta M^2).
\end{eqnarray}
\end{theorem}
This ends our discussion about the free energy in the ferromagnetic model.

Other properties of the model can be easily established. Introduce the 
Boltzmann-Gibbs state 
\begin{equation}\label{omega}
\omega_N(A)=Z_N^{-1}
\sum_{\sigma_1\dots\sigma_N} A \exp({1\over2} \beta N m^2)
\exp(\beta h \sum_{i}\sigma_i),
\end{equation}
where $A$ is any function of $\sigma_1\dots\sigma_N$.

The observable $m(\sigma)$ becomes self-averaging under 
$\omega_N$, in the infinite volume limit, in the sense that
\begin{equation}\label{selfaveraging}
\lim_{N\to\infty}\omega_N \bigl((m-M(\beta,h))^2\bigr)=0.
\end{equation}
This property of $m$ is the deep reason for the success of the strategy 
exploited before for the convergence of the free energy. Easy consequences 
are the following. In the infinite volume limit, the Boltzmann-Gibbs state 
becomes a factor state
\begin{equation}\label{factor}
\lim_{N\to\infty}\omega_N (\sigma_1\dots\sigma_s)=M(\beta,h)^s.
\end{equation}
A phase transition appears in the form of spontaneous magnetization. In 
fact, while for $h=0$ and $\beta\le1$ we have $M(\beta,h)=0$, on the other 
hand, for $\beta>1$, we have the discontinuity
\begin{equation}\label{+-}
\lim_{h\to0^{+}} M(\beta,h) = - \lim_{h\to0^{-}} M(\beta,h)\equiv 
M(\beta)>0.
\end{equation}
Fluctuations can also be easily controlled. In fact, one proves that the 
rescaled random variable $\sqrt{N}(m-M(\beta,h))$ tends in distribution, 
under $\omega_N$, to a centered Gaussian with variance given by the 
suscettibility
\begin{equation}\label{chi}
\chi(\beta,h)\equiv {\partial\over\partial h} M(\beta,h)\equiv
{{\beta(1-M^2)}\over{1- \beta(1-M^2)}}.
\end{equation}
Notice that the variance becomes infinite only at the critical point 
$h=0$, $\beta=1$, where $M=0$.

Now we are ready to attack the much more difficult spin glass model. But 
it will be surprising to see that, by following a simple extension of the 
methods here described, we will arrive to similar results.

\section{The basic definitions for the mean field spin glass model}

As in the ferromagnetic case, the generic configuration of the mean field spin 
glass model is defined 
through Ising spin variables
$\sigma_{i}=\pm 1$,  attached to each site $i=1,2,\dots,N$. 

But now there is an external quenched disorder 
given by the $N(N-1)/2$ independent and identical distributed random
variables $J_{ij}$, defined for each couple of sites. For the sake of simplicity,
we assume each $J_{ij}$ to be a centered
unit Gaussian with averages $E(J_{ij})=0$, $ E(J_{ij}^2)=1$. By quenched 
disorder we mean that the $J$ have a kind of stochastic external influence 
on the system, without partecipating to the thermal equilibrium.   

Now the Hamiltonian of the model, in some external field of strength $h$,  
is given by the mean field expression
\begin{equation}\label{Hsg}
H_N(\sigma,h,J)=-{1\over\sqrt{N}}\sum_{(i,j)}J_{ij}\sigma_i\sigma_j
-h\sum_{i}\sigma_i.
\end{equation}
Here, the first sum extends to all site couples, an the second to all sites. 
Notice the $\sqrt{N}$, necessary to ensure a good thermodynamic behavior to 
the free energy.

For a given inverse temperature $\beta$, let us now introduce the 
disorder dependent
partition function $Z_{N}(\beta,h,J)$ and
the quenched average of the free energy per site
$f_{N}(\beta,h)$,   
according to the  definitions
\begin{eqnarray}\label{Zsg}
&&Z_N(\beta,h,J)=\sum_{\sigma_1\dots\sigma_N}\exp(-\beta H_N(\sigma,h,J)),\\
\label{fsg}
&&-\beta f_N(\beta,h)=N^{-1} E\log Z_N(\beta,h,J). 
\end{eqnarray}
Notice that in (\ref{fsg}) the average $E$ with respect to the external noise is 
made \textit{after} the $\log$ is taken. This procedure is called quenched 
averaging. It represents the physical idea that the external noise does not 
partecipate to the thermal equilibrium. Only the $\sigma$'s are thermalized.

For the sake of simplicity, it is also convenient to write the partition
function in the following equivalent form. First of all let us introduce 
a family of centered Gaussian random variables ${\cal K}(\sigma)$, indexed by the  
configurations $\sigma$, and characterized by the covariances
\begin{equation}\label{cov}
E\bigl({\cal K}(\sigma){\cal K}(\sigma^\prime)\bigr)=q^2(\sigma,\sigma^\prime),
\end{equation}
where $q(\sigma,\sigma^\prime)$ are the overlaps between two generic 
configurations, defined by
\begin{equation}\label{overlap}
q(\sigma,\sigma^\prime)=
N^{-1}\sum_{i}\sigma_i\sigma^{\prime}_i,
\end{equation}
with the obvious bounds
$-1\le q(\sigma,\sigma^\prime)\le 1$, and the normalization 
$q(\sigma,\sigma) = 1$.
Then, starting from the definition (\ref{Hsg}), it is immediately seen that 
the partition function in (\ref{Zsg}) can be also written, by neglecting 
unessential constant terms, in the form  
\begin{equation}\label{Zsg'}     
Z_N(\beta,h,J)=
\sum_{\sigma_1\dots\sigma_N}\exp(\beta \sqrt{N\over2}{\cal K}(\sigma))
\exp(\beta h \sum_{i}\sigma_i),
\end{equation}    
which will be the starting point of our treatment.

\section{Basic formulae of derivation and interpolation}

We work in the following general setting. Let $U_i$ be a family
of centered Gaussian random variables, $i=1,\dots,K$, with covariance 
matrix given by $E(U_i U_j) \equiv S_{ij}$. We treat the index $i$ now as 
configuration space for some statistical mechanics system, with partition 
function $Z$ and quenched 
free energy given by
\begin{equation}\label{frU}
E\log\sum_i w_i \exp (\sqrt{t} U_i)\equiv E\log Z,
\end{equation}
where $w_i\ge0$ are generic weigths, and $t$ is a parameter ruling the 
strength of the interaction.

It would be hard to underestimate the relevance of the following derivation 
formula
\begin{eqnarray}
\nonumber
&& {d\over dt}E\log\sum_i w_i \exp (\sqrt{t} U_i)=\\
\nonumber
&& {1\over2}E(Z^{-1}\sum_i w_i \exp (\sqrt{t} U_i) S_{ii}\\
\label{derivation}
&& -{1\over2}
E(Z^{-2}\sum_i \sum_j w_i w_j \exp (\sqrt{t} U_i) \exp (\sqrt{t} U_j) 
S_{ij}).
\end{eqnarray}
The proof is straigthforward. Firstly we perform directly the $t$ 
derivative. Then, we notice that the random variables appear in expressions 
of the form $E(U_i F)$, were $F$ are functions of the $U$'s. These can be 
easily handled through the following integration by parts formula for generic 
Gaussian random variables, strongly reminiscent of the Wick theorem in 
quantum field theory,
\begin{equation}
\label{wick}
E(U_i F)=\sum_j S_{ij} E({\partial\over\partial{U_j}}F).
\end{equation}
Therefore, we see that always two derivatives are involved. The two terms 
in (\ref{derivation}) come from the action of the $U_j$ derivatives, the 
first acting on the Boltzmann factor, and giving rise to a Kronecker 
$\delta_{ij}$, the second acting on $Z^{-1}$, and giving rise to the minus 
sign and the duplication of variables. 

The derivation formula can be 
expressed in a more compact form by introducing replicas and suitable 
averages. In fact, let us introduce the state $\omega$ acting on 
functions $F$ of $i$ as follows
\begin{equation}\label{state}
\omega(F(i))=Z^{-1}\sum_i w_i \exp (\sqrt{t} U_i) F(i),
\end{equation}
together with the associated product state $\Omega$ acting on replicated 
configuration spaces $i_1,i_2,\dots,i_s$. By performing also a global $E$
average, finally we define the averages
\begin{equation}\label{bracket}
\med{F}_t\equiv E\Omega(F),
\end{equation}
where the subscript is introduced in order to recall the $t$ dependence of 
these averages.

Then, the equation (\ref{derivation}) can be written in a more compact form
\begin{equation}\label{derivation'}
{d\over dt}E\log\sum_i w_i \exp (\sqrt{t} U_i)=
{1\over2}\med{S_{i_1 i_1}}-{1\over2}
\med{
S_{i_1 i_2}}.
\end{equation}

Our basic comparison argument will be based on the following very simple 
theorem.
\begin{theorem}
\label{comparison}
Let $U_i$ and $\hat U_i$, for $i=1,\dots,K$, be independent families of 
centered Gaussian random variables, whose covariances satisfy the 
inequalities for generic configurations
\begin{equation}\label{dominance}
E(U_i U_j) \equiv S_{ij} \ge E(\hat U_i \hat U_j) \equiv \hat S_{ij},
\end{equation}
and the equalities along the diagonal
\begin{equation}\label{diagonal}
E(U_i U_i) \equiv S_{ii} = E(\hat U_i \hat U_i) \equiv \hat S_{ii},
\end{equation}
then for the quenched averages we have the inequality in the opposite sense
\begin{equation}\label{basic}
E\log \sum_i w_i\exp(U_i) \le E\log \sum_i w_i\exp(\hat U_i),
\end{equation} 
where the 
$w_i\ge 0$ are the same in the two expressions. 
\end{theorem}
Considerations of this kind are present in the mathematical 
literature of some years ago. Two typical references are \cite{Joag} 
and \cite{kahane}.

The proof is extremely simple and amounts to a straigthforward 
calculation. In fact, let us consider the interpolating expression
\begin{equation}\label{int}
E\log \sum_i w_i\exp(\sqrt{t}U_i + \sqrt{1-t}\hat U_i), 
\end{equation}
where $0\le t \le 1$. Clearly the two expressions under comparison 
correspond to the values $t=0$ and $t=1$ respectively. By taking the 
derivative with respect to $t$, with the help of the previous derivation 
formula, we arrive to the evaluation of the $t$ derivative in the form
\begin{eqnarray}
\nonumber
&& {d\over dt}E\log \sum_i w_i\exp(\sqrt{t}U_i + \sqrt{1-t}\hat U_i)=\\
\nonumber
&& {1\over2}E(Z^{-1}\sum_i w_i \exp (\sqrt{t} U_i) (S_{ii}-\hat S_{ii})\\
\label{derivation''}
&& -{1\over2}E(Z^{-2}\sum_i \sum_j w_i w_j \exp (\sqrt{t} U_i) \exp (\sqrt{t} U_j) 
(S_{ij}-\hat S_{ij}).
\end{eqnarray}
From the conditions assumed for the covariances, we immediately see that the 
interpolating 
function is nonincreasing in $t$, and the theorem follows.

The derivation formula and the comparison Theorem are not restricted to the 
Gaussian case. Generalizations in many directions are possible. For the 
diluted spin glass models and optimization problems we refer for example 
to \cite{FL},  and to 
\cite{LDS}, and references quoted there.

\section{The thermodynamic limit and the variational bounds} 

We give here some striking applications of the basic comparison Theorem.
In \cite{GTthermo} we have given a very simple proof of a long waited 
result, about the convergence of the free energy per site in the 
thermodynamic limit. Let us show the argument. Let us consider a system of 
size $N$ and two smaller systems of sizes $N_1$ and $N_2$ respectively, 
with $N=N_1+N_2$, as before in the ferromagnetic case. Let us now compare
\begin{equation}\label{ElogZ}     
E\log Z_N(\beta,h,J)=E\log
\sum_{\sigma_1\dots\sigma_N}\exp(\beta \sqrt{N\over2}{\cal K}(\sigma))
\exp(\beta h \sum_{i}\sigma_i),
\end{equation}
with
\begin{eqnarray}
\nonumber
&&E\log
\sum_{\sigma_1\dots\sigma_N}\exp(\beta 
\sqrt{N_1\over2}{\cal K}^{(1)}(\sigma^{(1)}))
\exp(\beta 
\sqrt{N_2\over2}{\cal K}^{(2)}(\sigma^{(2)}))
\exp(\beta h \sum_{i}\sigma_i) \\
\label{Z1Z2}
&&\equiv E\log Z_{N_1}(\beta,h,J)
+E\log Z_{N_2}(\beta,h,J) ,
\end{eqnarray}
where $\sigma^{(1)}$ are the $(\sigma_i,\ i=1,\dots,N_1)$, and
$\sigma^{(2)}$ are the $(\sigma_i,\ i=N_1+1,\dots,N)$. Covariances for
${\cal K}^{(1)}$ and ${\cal K}^{(2)}$ are expressed as in (\ref{cov}), but now the 
overlaps are substituted with the partial overlaps of the first and second 
block, $q_1$ and $q_2$ respectively. It is very simple to apply the comparison 
theorem. All one has to do is to observe that the obvious
\begin{equation}\label{q}
N q= N_1 q_1 + N_2 q_2,
\end{equation}
analogous to (\ref{m'}), implies, as in (\ref{m2}),
\begin{equation}\label{q2}
q^2 \le {N_1\over N} q_1^2 + {N_2\over N} q_2^2.
\end{equation}
Therefore, the comparison gives the superaddivity property, to be compared 
with (\ref{sub}),
\begin{equation}\label{super}
E\log Z_N(\beta,h,J)\ge E\log Z_{N_1}(\beta,h,J) + E\log Z_{N_2}(\beta,h,J).
\end{equation}
From the superaddivity property the existence of the limit follows in the 
form
\begin{equation}\label{limsg}
\lim_{N\to\infty}N^{-1}E\log Z_N(\beta,h,J)= \sup_{N}N^{-1}E\log Z_N(\beta,h,J),
\end{equation}
to be compared with (\ref{lim}).

The second application is in the form of the Aizenman-Sims-Starr 
generalized variational principle. Here, we will need to introduce some 
auxiliary system. The denumerable configuration space is given by the values of 
$\alpha=1,2,\dots$. We introduce also weights $w_{\alpha}\ge0$ 
for the $\alpha$ system, and suitably defined overlaps between two generic configurations 
$p(\alpha,\alpha^{\prime})$, with $p(\alpha,\alpha)=1$. 

A family of 
centered Gaussian random variables $\hat{{\cal K}}(\alpha)$, now indexed by the  
configurations $\alpha$, will be defined by the covariances
\begin{equation}\label{covhat}
E\bigl(\hat{{\cal K}}(\alpha) \hat{{\cal K}}(\alpha^\prime)\bigr)=
p^2(\alpha,\alpha^\prime).
\end{equation}  
We will need also a family of centered Gaussian random variables 
$\eta_{i}(\alpha)$, indexed by the sites $i$ of our original system and the 
configurations $\alpha$ of the auxiliary system, so that
\begin{equation}\label{eta}
E\bigl(\eta_i(\alpha) \eta_{i^\prime}(\alpha^\prime)\bigr)=
\delta_{i i^\prime} p(\alpha,\alpha^\prime).
\end{equation}

Both the probability measure $w_\alpha$, and the overlaps 
$p(\alpha,\alpha^{\prime})$ could depend on some additional external 
quenched noise, that does not appear explicitely in our notation.

In the following, we will denote by $E$ averages with respect to all 
random variables involved.

In order to start the comparison argument, we will consider firstly the 
case where the two  $\sigma$ and $\alpha$ systems are not coupled, so to 
appear factorized in the form
\begin{eqnarray}
\nonumber
&&E\log
\sum_{\sigma_1\dots\sigma_N}\sum_{\alpha} w_{\alpha}
\exp(\beta \sqrt{N\over2}{\cal K}(\sigma))
\exp(\beta \sqrt{N\over2}\hat{{\cal K}}(\alpha))
\exp(\beta h \sum_{i}\sigma_i) \\
\label{first}
&&\equiv E\log Z_N(\beta,h,J) +
E\log
\sum_{\alpha} w_{\alpha}
\exp(\beta \sqrt{N\over2}\hat{{\cal K}}(\alpha)).
\end{eqnarray}

In the second case the ${\cal K}$ fields are suppressed and the coupling 
between the two systems will be taken in a very simple form, by allowing 
the $\eta$ field to act as an external field on the $\sigma$ system. In 
this way the $\sigma$'s appear as factorized, and the sums can be 
explicitely performed. The chosen form for the second term in the 
comparison is
\begin{eqnarray}
\nonumber
&&E\log
\sum_{\sigma_1\dots\sigma_N}\sum_{\alpha} w_{\alpha}
\exp(\beta  \sum_{i}\eta_i(\alpha)\sigma_i)
\exp(\beta h \sum_{i}\sigma_i)\\
\label{second}
&& \equiv N\log2 +
E\log \sum_{\alpha} w_{\alpha}(c_1 c_2 \dots c_N), 
\end{eqnarray}
where we have defined
\begin{equation}\label{ci}
c_i = \cosh \beta(h+\eta_i(\alpha)),
\end{equation}
as arising from the sums over $\sigma$'s.

Now we apply the comparison Theorem. In the first case, the covariances 
involve the sums of squares of overlaps
\begin{equation}\label{squares}
{1\over2} \bigl(q^2(\sigma,\sigma^\prime)+p^2(\alpha,\alpha^\prime)\bigr). 
\end{equation}
In the second case, a very simple calculation shows that the covariances 
involve the overlap products 
\begin{equation}\label{product}
q(\sigma,\sigma^\prime) p(\alpha,\alpha^\prime\prime). 
\end{equation}
Therefore, the comparison is very easy and, by collecting all expressions, 
we end up with the useful estimate, as in \cite{ASS}, holding for any 
auxiliary system as defined before,
\begin{eqnarray}
\label{variational} 
&&N^{-1} E\log Z_N(\beta,h,J) \le\\
\nonumber
&&\log2 + N^{-1} \bigl(E\log \sum_{\alpha} w_{\alpha}(c_1 c_2 \dots c_N)-
E\log
\sum_{\alpha} w_{\alpha}
\exp(\beta \sqrt{N\over2}\hat{{\cal K}}(\alpha))\bigr).
\end{eqnarray}

\section{The Parisi representation for the free energy}

We refer to the original paper \cite{P}, and to the extensive 
review given in \cite{MPV}, for the general motivations, and the 
derivation of the broken replica  {\it Ansatz}, in the frame of the 
ingenious replica trick. Here we limit ourselves to a synthetic 
description of its general structure, independently from the replica trick

First of all, let us introduce the convex space ${\cal X}$ of the functional 
order parameters $x$, as nondecreasing functions of the auxiliary variable 
$q$,
both $x$ and $q$ taking
values on the interval $[0,1]$, {\it i.e.}
\begin{equation}
\label{x}
{\cal X}\ni x : [0,1]\ni q \rightarrow x(q) \in [0,1].
\end{equation}
Notice that we call $x$ the function, and $x(q)$ its values.
We introduce a metric on ${\cal X}$ through the $L^{1}([0,1], dq)$ norm, where 
$dq$ is the Lebesgue measure.

For our purposes, we will consider the case of piecewise constant functional order 
parameters, characterized by an integer $K$, and two sequences $q_0, q_1, 
\dots, q_K$, $m_1, m_2, \dots, m_K$ of numbers satisfying
\begin{equation}
\label{qm}
0=q_0\le q_1 \le \dots \le q_{K-1} \le q_K=1,\,\,\, 0\le m_1 \le m_2 \le \dots 
\le m_K \le 1,
\end{equation}
such that
\begin{eqnarray}
\nonumber
x(q)=m_1 \,\,\mbox{for}\,\, 0=q_0\le q < q_1,\,\,\, 
x(q)=m_2 &\mbox{for}& q_1\le q < q_2,\\
\label{xpcws}
\ldots,
 x(q)=m_K &\mbox{for}& q_{K-1}\le q \le q_K.
\end{eqnarray}
In the following, we will find convenient to define also $m_0\equiv 0$, 
and $m_{K+1}\equiv 1$. The replica symmetric case of Sherrington and 
Kirkpatrick corresponds to 
\begin{equation}
\label{replicas}
K=2,\,\, 
q_1=\bar q, \,\, m_1=0,\,\, m_2=1.
\end{equation}

Let us now introduce the function $f$, with values $f(q,y;x,\beta)$, of 
the variables $q\in[0,1]$, $y\in R$, depending also on the functional order 
parameter $x$, and on the inverse temperature $\beta$, defined as the 
solution of the nonlinear antiparabolic equation 
\begin{equation}
\label{antipara}
(\partial_q f)(q,y)+
{1\over2}(\partial_y^2 f)(q,y)+{1\over2}x(q)({\partial_y f})^2(q,y)=0,
\end{equation}
with final condition
\begin{equation}
\label{final}
f(1,y)=\log\cosh(\beta y).
\end{equation}
Here, we have stressed only the dependence of $f$ on $q$ and $y$.

It is very simple to integrate Eq.~(\ref{antipara}) when $x$ is piecewise 
constant. In fact, consider $x(q)=m_a$, for $q_{a-1}\le q \le q_{a}$, 
firstly with $m_a>0$. Then, 
it is immediately seen that the correct solution of Eq.~(\ref{antipara}) in 
this interval, with the right final boundary condition at $q = q_{a}$, is 
given by
\begin{equation}
\label{solution}
f(q,y)=\frac{1}{m_a}\log \int\exp\bigl({m_{a} f(q_a,y+z\sqrt{q_a-q})}\bigr)\,d\mu(z),
\end{equation}
where $d\mu(z)$ is the centered unit Gaussian measure on the real line. On 
the other hand, if $m_a=0$, then (\ref{antipara}) loses the nonlinear part 
and the solution is given by
\begin{equation}
\label{solution0}
f(q,y)= \int f(q_a,y+z\sqrt{q_a-q})\,d\mu(z),
\end{equation}
which can be seen also as deriving from (\ref{solution}) in the limit $m_a \to 0$.
Starting from the last interval $K$, and using (\ref{solution}) iteratively on 
each interval, we easily get the solution of (\ref{antipara}), 
(\ref{final}), in the case of piecewise order parameter $x$, as in
(\ref{xpcws}), through a chain of interconnected Gaussian integrations.

Now we introduce the following important definitions.
The trial auxiliary function, associated to a given mean field spin glass 
system, as described in Section 3, depending on the functional order 
parameter $x$, is defined as
\begin{equation}
\label{trial}
\log 2 + 
f(0,h;x,\beta)-\frac{\beta^2}{2}\int_{0}^{1} 
q\, x(q)\,dq.
\end{equation}
Notice that in this expression the function $f$ appears evaluated at $q=0$, 
and $y=h$, where $h$ is the value of the external magnetic field. This 
trial expression shoul be considered as the analog of that appearing in 
(\ref{lb}) for the ferromagnetic case.

The Parisi spontaneously broken replica symmetry expression for the free 
energy is given by the definition
\begin{equation}\label{parisi}
-\beta f_P(\beta,h) \equiv \inf_x \bigl(\log 2 + 
f(0,h;x,\beta)-\frac{\beta^2}{2}\int_{0}^{1} 
q\, x(q)\,dq \bigr),
\end{equation}
where the infimum is taken with respect to all functional order parameters 
$x$. 
Notice that the infimum appears here, as compared to the supremum in the 
ferromagnetic case.

In \cite{Grepli}, by exploiting a kind of generalized comparison argument, 
involving a suitably defined interpolation function, we have established 
the following important result.

\begin{theorem}
\label{main}
For all values of the inverse temperature $\beta$, and the external 
magnetic field $h$, and for any functional order parameter $x$, the 
following bound holds
$$
N^{-1}E\log Z_{N}(\beta,h,J)\le \log 2 + 
f(0,h;x,\beta)-\frac{\beta^2}{2}\int_{0}^{1} 
q\, x(q)\,dq ,
$$
uniformly in $N$. Consequently, we have also
$$
N^{-1}E\log Z_{N}(\beta,h,J)\le\inf_x \bigl(\log 2 + 
f(0,h;x,\beta)-\frac{\beta^2}{2}\int_{0}^{1} 
q\, x(q)\,dq \bigr),
$$
uniformly in $N$.
\end{theorem}

However, this result can be understood also in the frame of the 
generalized variational principle established by Aizenman-Sims-Starr and 
described before.

In fact, one can easily show that there exist an $\alpha$ systems such that
\begin{equation}
\label{c1c2}
N^{-1}E\log \sum_{\alpha} w_{\alpha}c_1 c_2 \dots c_N\equiv f(0,h;x,\beta),
\end{equation}
\begin{equation}
\label{Kappa}
N^{-1}E\log
\sum_{\alpha} w_{\alpha}
\exp(\beta \sqrt{N\over2}\hat{{\cal K}}(\alpha))\equiv\frac{\beta^2}{2}\int_{0}^{1} 
q\, x(q)\,dq ,
\end{equation}
uniformly in $N$. This result stems from previous work of Derrida, Ruelle, 
Neveu,  
Bolthausen, Sznitman, Aizenman, Contucci, Talagrand, Bovier, and others, and in a sense is 
implicit in the treatment given in \cite{MPV}. It can be reached in a very 
simple way. Let us sketch the argument.

First of all, let us consider the Poisson point process $y_1\ge y_2 \ge 
y_3 \dots$, uniquely characterized by the following conditions. For any 
interval $A$, introduce the occupation 
numbers $N(A)$, defined by
\begin{equation}
\label{NA}
N(A)=\sum_\alpha \chi (y_\alpha \in A),
\end{equation}
where $\chi()=1$, if the random variable $y_\alpha$ belongs to the 
interval $A$, and $\chi()=0$, otherwise. We assume that $N(A)$ and $N(B)$ 
are independent if the intervals $A$ and $B$ are disjoint, and moreover 
that for each $A$, the random variable $N(A)$ has a Poisson distribution 
with parameter
\begin{equation}
\label{muA}
\mu(A)=\int_a^b \exp(-y)\ dy,
\end{equation}
if $A$ is the interval $(a,b)$, \textit{i.e.}
\begin{equation}
\label{PNA}
P(N(A)=k)=\exp(-\mu(A)) \mu(A)^k /k!.
\end{equation}
We will exploit $-y_\alpha$ as energy levels for a statistical mechanics 
systems with configurations indexed by $\alpha$. For a parameter $0<m<1$, 
playing the role of inverse temperature, we can introduce the partition 
function 
\begin{equation}
\label{v}
v=\sum_\alpha \exp({y_\alpha\over m}).
\end{equation}
 For $m$ in the given interval 
it turns out that $v$ is a very well defined random variable, with the sum 
over $\alpha$ extending to infinity. In fact, there is a strong inbuilt 
smooth cutoff in the very definition of the stochastic energy levels. 

From 
the general properties of Poisson point processes it is very well known that 
the following basic invariance property holds. Introduce a random 
variable $b$, independent of $y$, subject to the condition $E(\exp b)=1$, 
and let $b_\alpha$ be independent copies. Then, the randomly biased point process 
$y^{\prime}_\alpha=y_\alpha + b_\alpha,\ \ \alpha=1,2,\dots$ is 
equivalent to the original one in distribution. An immediate consequence 
is the following. Let $f$ be a random variable, independent of $y$, such 
that $E(\exp f)<\infty$, and let $f_\alpha$ be independent copies. Then 
the two random variables
\begin{equation}
\label{yf}
\sum_\alpha \exp({y_\alpha\over m}) \exp(f_\alpha),
\end{equation}   
\begin{equation}
\label{yexp}\sum_\alpha \exp({y_\alpha\over m}) E(\exp(mf))^{1\over m}
\end{equation}
have the same distribution. In particular they can be freely substituted 
under averages.

The auxiliary system which gives rise to the Parisi representation 
according to (\ref{c1c2}) (\ref{Kappa}), for a piecewise constant order 
parameter, is expressed in the following way. Now $\alpha$ will be a 
multi-index $\alpha=(\alpha_1,\alpha_2,\dots,\alpha_K)$, where each 
$\alpha_a$ runs on $1,2,3,\dots$. Define the Poisson point process 
$y_{\alpha_1}$, then, independently, for each value of $\alpha_1$ 
processes $y_{\alpha_1 \alpha_2}$, and so on up to $y_{\alpha_1 
\alpha_2\dots\alpha_K}$. Notice that in the cascade of independent 
processes $y_{\alpha_1}, y_{\alpha_1 \alpha_2}, \dots y_{\alpha_1 
\alpha_2\dots\alpha_K}$, the last index refers to the numbering of the 
various points of the process, while the first indexes denotes independent 
copies labelled by the corresponding $\alpha$'s.

The weights $w_\alpha$ have to be chosen according to the definition
\begin{equation}
\label{w}
w_\alpha=\exp{y_{\alpha_1}\over m_1} \exp{y_{\alpha_1 \alpha_2}\over 
m_2}\dots \exp{y_{\alpha_1 \alpha_2\dots\alpha_K}\over m_K}.
\end{equation}
The cavity fields $\eta$ and ${\cal K}$ have the following expression in 
terms of independent unit Gaussian random variables 
$
J^i_{\alpha_1},J^i_{\alpha_1 \alpha_2},\dots,J^i_{\alpha_1 
\alpha_2\dots\alpha_K}$,
$J^{\prime}_{\alpha_1},J^{\prime}_{\alpha_1 \alpha_2},\dots,J^{\prime}_{\alpha_1 
\alpha_2\dots\alpha_K}$,
\begin{equation}
\label{eta'}
\eta_i(\alpha)=\sqrt{q_1-q_0}J^i_{\alpha_1}+\sqrt{q_2-q_1}J^i_{\alpha_1 \alpha_2}
+\dots +\sqrt{q_K-q_{K-1}}J^i_{\alpha_1 
\alpha_2\dots\alpha_K},
\end{equation}
\begin{equation}
\label{Kappa'}
{\cal K}(\alpha)=\sqrt{q_1^2-q_0^2}J^{\prime}_{\alpha_1}+
\sqrt{q_2^2-q_1^2}J^{\prime}_{\alpha_1 \alpha_2}
+\dots +\sqrt{q_K^2-q_{K-1}^2}J^{\prime}_{\alpha_1 
\alpha_2\dots\alpha_K}.
\end{equation}
It is immediate to verify that 
$E(\eta_i(\alpha)\eta_{i^{\prime}}(\alpha^{\prime})$ is zero if 
$i\ne i^{\prime}$, while
\begin{eqnarray}
\nonumber
E(\eta_i(\alpha)\eta_{i}(\alpha^{\prime}))&=& 0\ \mbox{if}\ 
\alpha_1\ne\alpha_1^{\prime},\\
\nonumber
&=&q_1\ \mbox{if}\ \alpha_1=\alpha_1^{\prime},\alpha_2\ne\alpha_2^{\prime},\\
\nonumber
&=&q_2\ \mbox{if}\ \alpha_1=\alpha_1^{\prime},\alpha_2=\alpha_2^{\prime},
\alpha_3\ne\alpha_3^{\prime},\\
\nonumber
&\dots&\\
\label{etaeta}
&=&1\ \mbox{if}\ \alpha_1=\alpha_1^{\prime},\alpha_2=\alpha_2^{\prime},\dots,
\alpha_K=\alpha_K^{\prime}.  
\end{eqnarray}
Similarly, we have
\begin{eqnarray}
\nonumber
E({\cal K}(\alpha){\cal K}(\alpha^{\prime}))&=& 0\ \mbox{if}\ 
\alpha_1\ne\alpha_1^{\prime},\\
\nonumber
&=&q_1^2\ \mbox{if}\ \alpha_1=\alpha_1^{\prime},\alpha_2\ne\alpha_2^{\prime},\\
\nonumber
&=&q_2^2\ \mbox{if}\ \alpha_1=\alpha_1^{\prime},\alpha_2=\alpha_2^{\prime},
\alpha_3\ne\alpha_3^{\prime},\\
\nonumber
&\dots&\\
\label{KappaKappa}
&=&1\ \mbox{if}\ \alpha_1=\alpha_1^{\prime},\alpha_2=\alpha_2^{\prime},\dots,
\alpha_K=\alpha_K^{\prime}.  
\end{eqnarray}
This ends the definition of the $\alpha$ system, associated to a given 
piecewise constant order parameter.

Now, it is simple to verify that (\ref{c1c2}) (\ref{Kappa}) hold.
Let us consider for example (\ref{c1c2}). With the $\alpha$ system chosen 
as before, the repeated application of the stochastic equivalence of 
(\ref{yf}) and (\ref{yexp}) will give rise to a sequence of interchained 
Gaussian integrations exactly equivalent to those arising from the 
expression for $f$, as solution of the equation (\ref{antipara}). For 
(\ref{yexp}), there are equivalent considerations.  

Therefore we see that the estimate in Theorem~\ref{main} is also 
a consequence of 
the generalized variational principle.

Up to this point we have seen how to obtain upper bounds. The problem 
arises whether, as in the ferromagnetic case, we can also get lower 
bounds, so to shrink the thermodynamic limit to the value given by the 
$\inf_x$ in Theorem~\ref{main}. After a short announcement in \cite{Tb}, 
Michel Talagrand wrote an extended paper \cite{Topus}, 
to appear on Annals of Mathematics, where the complete proof of the control 
of the lower bound is firmly established. We refer to the original paper 
for the complete details of this remarkable achievement. About the 
methods, here we only recall that in \cite{Grepli} we have given also the 
corrections to the bounds appearing in Theorem~\ref{main}, albeit in a 
quite complicated form. Talagrand has been able to establish that these 
corrections do in 
fact vanish in the thermodynamic limit.

In conclusion, we can establish the following extension of Theorem~\ref{tlim} 
to spin glasses.
\begin{theorem}
\label{tlimsg}
For the mean field spin glass model we have
\begin{eqnarray}
&&\lim_{N\to\infty}N^{-1}E\log Z_N(\beta,h,J) =  \sup_N N^{-1}E\log 
Z_N(\beta,h,J)\\
&&=\inf_x \bigl(\log 2 + 
f(0,h;x,\beta)-\frac{\beta^2}{2}\int_{0}^{1} 
q\, x(q)\,dq \bigr).
\end{eqnarray}
\end{theorem}

\section{Diluted models}

Diluted models, in a sense, play a role intermediate between the mean field 
case and the short range case. In fact, while in the mean field model each 
site is interacting with all other sites, on the other hand, in the 
diluted model, each site is interacting with only a fixed number of other 
sites. However, while for the short range models there is a definition of 
distance among sites, relevant for the interaction, no such definition 
appears in the diluted models, where all sites are in any case equivalent. 
From this point of view, the diluted models are structurally similar to the 
mean field models, and most of the techniques and results explained before 
can be extended to them.

Let us define a typical diluted model. The quenched noise is described as 
follows. Let $K$ be a Poisson random variable 
with parameter $\alpha N$, where $N$ is the number of sites, and $\alpha$ 
is a parameter entering the theory, together with the temperature. We 
consider also a sequence of independent centered random variables 
$J_1,J_2,\dots$, and a sequence of discrete independent 
random 
variables $i_1,j_1,i_2,j_2,\dots$, uniformly distributed over the set of 
sites $1,2,\dots,N$. Then we assume as Hamiltonian
\begin{equation}\label{Hdiluted}
 H_N (\sigma) = -\sum_{k=0}^K J_k \sigma_{i_k} \sigma_{j_k}.  
\end{equation}
Only the variables $\sigma$ partecipate to thermodynamic equilibrium. All 
noise coming from $K,J_k,i_k,j_k$ is considered quenched, and it is 
not explicitely indicated in our notation for $H$.

The role played by Gaussian integration by parts in the 
Sherrington-Kirckpatrick model, here is assumed by the following elementary 
derivation formula, holding for Poisson distributions,
\begin{eqnarray}
\nonumber
{d\over dt} P(K=k,t\alpha N)&\equiv & 
{d\over dt} \exp(-t\alpha N) (t\alpha 
N)^k/{k!} \\
\label{dtPoisson}
&= &\alpha N (P(K=k-1,t\alpha N)-P(K=k,t\alpha N)).  
\end{eqnarray}
Then, all machinery of interpolation can be easily extended to the diluted 
models, as firstly recognized by Franz and Leone in \cite{FL}.

In this way, the superaddivity property, the thermodynamic limit, and the 
generalized variational principle can be easily established. We refer to 
\cite{FL}, and \cite{LDS}, for a complete treatment.

There is an important open problem here. While in the fully connected case 
the Poisson probability cascades provide the rigth auxiliary $\alpha$ 
systems to be exploited in the variational principle, on the other hand in 
the diluted case more complicated probability cascades have been proposed, 
as shown for example in \cite{FL}, and in \cite{PT}. On the other hand, in 
\cite{LDS}, the very interesting proposal has been made that also in the 
case of diluted models the Poisson probability cascades play a very 
important role. Of course here the way how the auxiliary system interact 
with the original system is different, and involves a multi-overlap 
structure as explained in \cite{LDS}. In this way a kind of very deep 
universality is emerging. Poisson probability cascades are a kind of 
universal class of auxiliary systems. The different models require 
different cavity fields ruling the interaction between the original system 
and the auxiliary system. But further work will be necessary in order to 
clarify this very important issue. For results about diluted models in the 
high temperature region, we refer to \cite{GTdiluted}.

\section{The short range model and its connections with the mean field 
version}

The investigation of the connections between the short range version of the 
model and its mean field version are at the beginning. Here we limit 
ourselves to a synthetic description of what should be done, and to a 
short presentation of the results obtained so far.

First of all, according to the conventional wisdom, 
the mean field version should be a kind of limit of the short range model 
on a lattice in dimension $d$, when $d\to\infty$, with a proper rescaling 
of the strength of the Hamiltonian, of the form $d^{- {1\over2}}$. Results of 
this kind are very well known in the ferromagnetic case, but the present 
technology of interpolation does not seem sufficient to assure a proof in 
the spin glass case. So this very basic result is still missing. In 
analogy with the ferromagnetic case, it would be necessary to arrive at the 
notion of a critical dimension, beyond which the features of the mean 
field case still hold, for example in the expression of the critical 
exponents and in the ultrametric hierarchical structure of the pure phases, 
or at least for the overlap distributions. For physical dimensions less 
than the critical one, then the short range model would need corrections 
with respect to its mean field version. Therefore, this is a completely 
open problem.

Moreover, always according to the conventional wisdom, 
the mean field version should be a kind of limit of the short range models, 
in finite fixed dimensions, as the range of the interaction goes to 
infinity, with proper rescaling. Important work of Franz and Toninelli 
shows that this is effectively the case, if a properly defined Kac limit 
is performed. Here, interpolation methods are effective, and we refer to 
\cite{GTkac}, \cite{FTletter}, \cite{FT} for full details.

Due to the lack of efficient analytical methods, it is clear that 
numerical simulations play a very important role in the study of the 
physical properties emerging from short range spin glass models. In 
particular, we refer to \cite{MPR} and \cite{MPRZ}, for a detailed account 
about the evidence, coming from theoretical considerations and extensive 
computer simulations, that some of the more relevant features of the 
spontaneous replica breaking scheme of the mean field are also present in 
short range models in three dimensions. Different views are expressed for 
example in \cite{NS}, where it is argued that the phase space structure of 
short range spin glass models is much simpler than that foreseen by the 
Parisi spontaneous replica symmetry mechanism.

Such very different views, both apparently strongly supported by reasonable 
theoretical considerations 
and powerful numerical simulations, are a natural consequences of the 
extraordinary difficulty of the problem.

It is clear that extensive additional work will be necessary before the 
clarification of the physical features exhibited by the realistic short 
range spin glass models.   

\section{Conclusion and outlook for future developments}

As we have seen, in these last few years there has been an impressive progress in the 
understanding of 
the mathematical structure of spin glass models, mainly due to the 
systematic exploration of comparison and interpolation methods. 
However many important problems are still open. The most important one is 
to establish rigorously the full hierarchical ultrametric organization of 
the overlap distributions, as appears in Parisi theory, 
and to fully understand the decomposition in pure states of the glassy phase, at 
low temperatures.

Moreover, is would be important to extend these methods to other important 
disordered models as for example neural networks. Here the difficulty is 
that the positivity arguments, so essential in comparison methods, do not 
seem to emerge naturally inside the structure of the theory.

Finally, the problem of connecting properties of the short range model, 
with those arising in the mean field case, is still almost completely open.

\vspace{.5cm}
{\bf Acknowledgments}

We gratefully acknowledge useful conversations with Michael Aizenman, 
Pierluigi Contucci,
Giorgio Parisi and Michel Talagrand. The strategy explained in this report 
grew out from a 
systematic exploration of comparison and interpolation methods, developed in 
collaboration with Fabio Lucio Toninelli, and Luca De Sanctis.

This work was supported in part by MIUR 
(Italian Minister of Instruction, University and Research), 
and by INFN (Italian National Institute for Nuclear Physics).

\end{document}